\documentclass[reprint,amsmath,amssymb,aps,pra]{revtex4-1}

\usepackage{graphicx}
\usepackage{dcolumn}
\usepackage{bm}
\usepackage{color, soul}
\sethlcolor{cyan}

\usepackage[table]{xcolor}
\usepackage{wasysym}
\usepackage{subfigure}
\usepackage{commath}
\usepackage{braket}
\usepackage{amsmath}
\usepackage{placeins}
\let\oldhat\hat
\renewcommand{\hat}[1]{\oldhat{\mathbf{#1}}}
\usepackage{xcolor}

\usepackage{amsmath}
\usepackage{amsfonts}
\usepackage{amssymb}
\usepackage{amsthm}

\usepackage{siunitx}
\usepackage{booktabs}
\usepackage{multirow}
\usepackage{epstopdf}
\usepackage{longtable}
\usepackage{array}
\usepackage{latexsym}
\usepackage[latin1]{inputenc}
\usepackage{textcomp}
\usepackage[nice]{nicefrac}

\begin{document}

\preprint{APS/123-QED}

\title{Isotropic Light vs Six-Beam Molasses for Doppler Cooling of Atoms From Background Vapor - Theoretical Comparison}

\author{St\'ephane Tr\'emine}
\email[E-mail: ]{stephane.tremine@gmail.com}
\affiliation{LNE-SYRTE, Observatoire de Paris, PSL Research University, CNRS, Sorbonne Universités, UPMC Univ. Paris 06, 61 Avenue de l'Observatoire, 75014 Paris, France}
\author{Emeric de Clercq}
\affiliation{LNE-SYRTE, Observatoire de Paris, PSL Research University, CNRS, Sorbonne Universités, UPMC Univ. Paris 06, 61 Avenue de l'Observatoire, 75014 Paris, France}
\author{Philippe Verkerk}
\affiliation{Laboratoire de Physique des Lasers Atomes et Mol\'ecules, CNRS, Universit\'e Lille 1 , 59655 Villeneuve d'Ascq, France}
\date{\today}

\begin{abstract}
We present a 3D theoretical comparison between the radiation-pressure forces exerted on an atom in an isotropic light cooling scheme and in a six-beam molasses. We demonstrate that, in the case of a background vapor where all the space directions of the atomic motion have to be considered, the mean cooling rate is equal in both configurations. Nevertheless, we also point out what mainly differentiates the two cooling techniques: the force component orthogonal to the atomic motion. If this transverse force is always null in the isotropic light case, it can exceed the radiation-pressure-force longitudinal component in the six-beam molasses configuration for high atomic velocities, hence reducing the velocity capture range.
\end{abstract}

\maketitle
\section{\label{sec:level1}Introduction}

Laser cooling of neutral atoms in red-shifted monochromatic isotropic light has initially been proposed by Wang in $1979$~\cite{Wang1979} to decelerate atoms from a thermal beam, and then demonstrated experimentally by Ketterle \textit{et al.} in $1992$~\cite{Ketterle1992} and Batelaan \textit{et al.} in $1994$~\cite{Batelaan1994}. Both experiments consist in a thermal beam passing through a tube made in spectralon$^{\circledR}$~\cite{Labsphere} (a material with a diffuse reflectivity $R\sim 99\%$) into which laser light was coupled using multimode optical fibers in order to build the required isotropic field. This cooling technique has also been demonstrated by Wang \textit{et al.} in $1994$~\cite{Wang1994,Chen1994} using a thermal beam going through an integrating sphere made in copper, with a diffusive MgO coating on the inner surface of the cavity. This atom-slowing scheme was an elegant alternative to Zeeman slowing~\cite{Prodan1982}, chirped slowing~\cite{Balykin1980} or white-light slowing~\cite{Moi1984,Hoffnagle1988} used to compensate the Doppler shift variation as the atoms decelerate, in order to maintain a maximum effectiveness of the cooling forces.\\

The authors of references~\cite{Ketterle1992,Wang1994} also suggested that isotropic light cooling (ILC) could be used to produce optical molasses as an attractive alternative to the celebrated technique using three orthogonal pairs of counterpropagating collimated laser beams~\cite{Chu1985}. In this case, and when applied on a background gas as in reference~\cite{Monroe1990}, ILC was even expected by Ketterle \textit{et al.} to be more effective than the standard six-beam configuration \cite{Ketterle1992}. This expectation was based on a theoretical comparison of both configurations, showing that the cooling rate in isotropic light should be higher than in a six-beam molasses (SBM) for atomic velocities $v>\Gamma/k$ ($\Gamma$ and $k$ being the natural linewidth of the atomic transition implied in the cooling process and the wave vector norm of the cooling light, respectively), when calculated for the same laser light detuning $\Delta$ and photon density.\\

ILC of an atomic vapor was first demonstrated by Aucouturier \textit{et al} in $1997$~\cite{Aucouturier1997} by enclosing a vapor cell in a cavity made in spectralon$^{\circledR}$, and then using cavities made in copper with polished inner surfaces reaching reflective coefficients as high as $96\%$~\cite{Guillot2001}. This cooling technique allows the conception of compact and  robust sources of cold atoms for spectroscopy applications or sensors. Some high performances atomic clocks based on ILC are already in development~\cite{Tremine2005,Esnault2010,Liu2015,Liu2016}.\\

Here, we pursue, in the case of a background atomic gas, the theoretical comparison initiated by Ketterle \textit{et al.} between collimated-beam and isotropic-light three-dimensional laser cooling \cite{Ketterle1992}. The cooling radiation-pressure force expressions in both configurations are briefly reminded in Section~\ref{sec:3DRadiationPressureForcesExpressions} in the context of the two-level atom model, while considering plane waves and low laser intensities. We distinguish the longitudinal force, along the atomic motion, and the transverse force, perpendicular to it. The magnitude of both components, and their mean values (averaged over all space directions), are addressed in Section~\ref{sec:TheoreticalComparison} for both configurations. The computations are performed in the case of Cs atom. In this paper, our aim is to point out the specificities of each cooling scheme in the case of a background vapor, in which all the directions of the atomic velocity vector $\bm{\mathrm{v}}$ are represented, as opposed to the thermal beam case.

\section{\label{sec:3DRadiationPressureForcesExpressions}3D radiation-pressure force expressions}

According to the Doppler cooling theory~\cite{Cook1980}, the average radiation-pressure force exerted on a two-level atom irradiated by a laser beam that is assumed to be a plane wave of angular frequency $\omega$ and wavelength $\lambda$ is given by
\begin{eqnarray}
\bm{\mathrm{\mathcal{F}}}=\frac{\Gamma}{2}\frac{s}{1+s}\hbar \bm{\mathrm{k}},
\label{eq:F1Faisceau}
\end{eqnarray}
where $\bm{\mathrm{k}}$ is the wave vector related to the plane wave ($k=2\pi/\lambda$), and $\hbar$ is the reduced Planck's constant. $s$ is the generalized saturation parameter, taking into account the Doppler shift, defined by
\begin{eqnarray}
s=\frac{s_{0}}{1+4(\Delta - \bm{\mathrm{k}}\cdot\bm{\mathrm{v}})^{2}/\Gamma^{2}} \hspace{0.3cm} \text{with} \hspace{0.3cm} s_{0}=\frac{2\Omega_{R}^{2}}{\Gamma^{2}}=\frac{I}{I_{s}},
\label{eq:SaturationParameter}
\end{eqnarray}
$s_0$ being the on-resonance saturation parameter, with $\Delta=\omega-\omega_{0}$ the detuning of the light-wave frequency $\omega$ relatively to the atomic transition frequency $\omega_{0}$; $\Omega_{R}$ the Rabi frequency associated to the laser intensity $I$; and $I_{s}$ the saturation intensity related to the atomic transition involved in the cooling process. It is worth to note that the radiation-pressure force (\ref{eq:F1Faisceau}) is always in the direction of propagation of the light, and is maximum when the laser detuning counterbalances the Doppler shift such as:
\begin{align}
\Delta=\omega-\omega_{0}=\bm{\mathrm{k}}\cdot\bm{\mathrm{v}}.
\label{eq:ResonanceCondition}
\end{align}
Due to the stochastic nature of the spontaneous emission process that follows each photon absorption, the associated recoil suffered by the atom will be considered to be null on average, assuming that enough absorption-emission cycles occur during the atom-light interaction. The diffusion of the atom momentum will not be tackled here.

    \subsection{\label{subsec:6BeamMolassesConfiguration}Six-beam configuration}

    We now consider the case of two identical but counterpropagating laser beams, along the $x$ axis, of wave vectors $\bm{\mathrm{k_x}}$ and $-\bm{\mathrm{k_x}}$. We suppose equal intensities and an optically thin medium, so that the laser intensities can be considered constant.  Assuming that the two waves act independently on the atoms (which is only valid at low saturation parameters $s_{0}\leq 1$), then the average total radiation-pressure force, collinear to the laser beams, is simply given by the addition of both separate forces~\cite{Cook1980}:
    \begin{eqnarray}
     \bm{\mathrm{\mathcal{F}^{1\mathrm{D}}_{x}}}=\mathcal{F}^{+}_x \hat{e}_x +\mathcal{F}^{-}_x \hat{e}_x,
    \label{eq:Fcol1D}
    \end{eqnarray}
    where $\mathcal{F}^{\pm}_x$ refers to the force (\ref{eq:F1Faisceau}) for the $\pm \bm{\mathrm{k_x}}$ wave, and $\hat{e}_x$ is the unit vector of the $x$ axis.\\

\begin{figure}[htb]
    \centering
    \includegraphics[trim=0mm 0mm 0mm 0mm, width=0.6\linewidth]{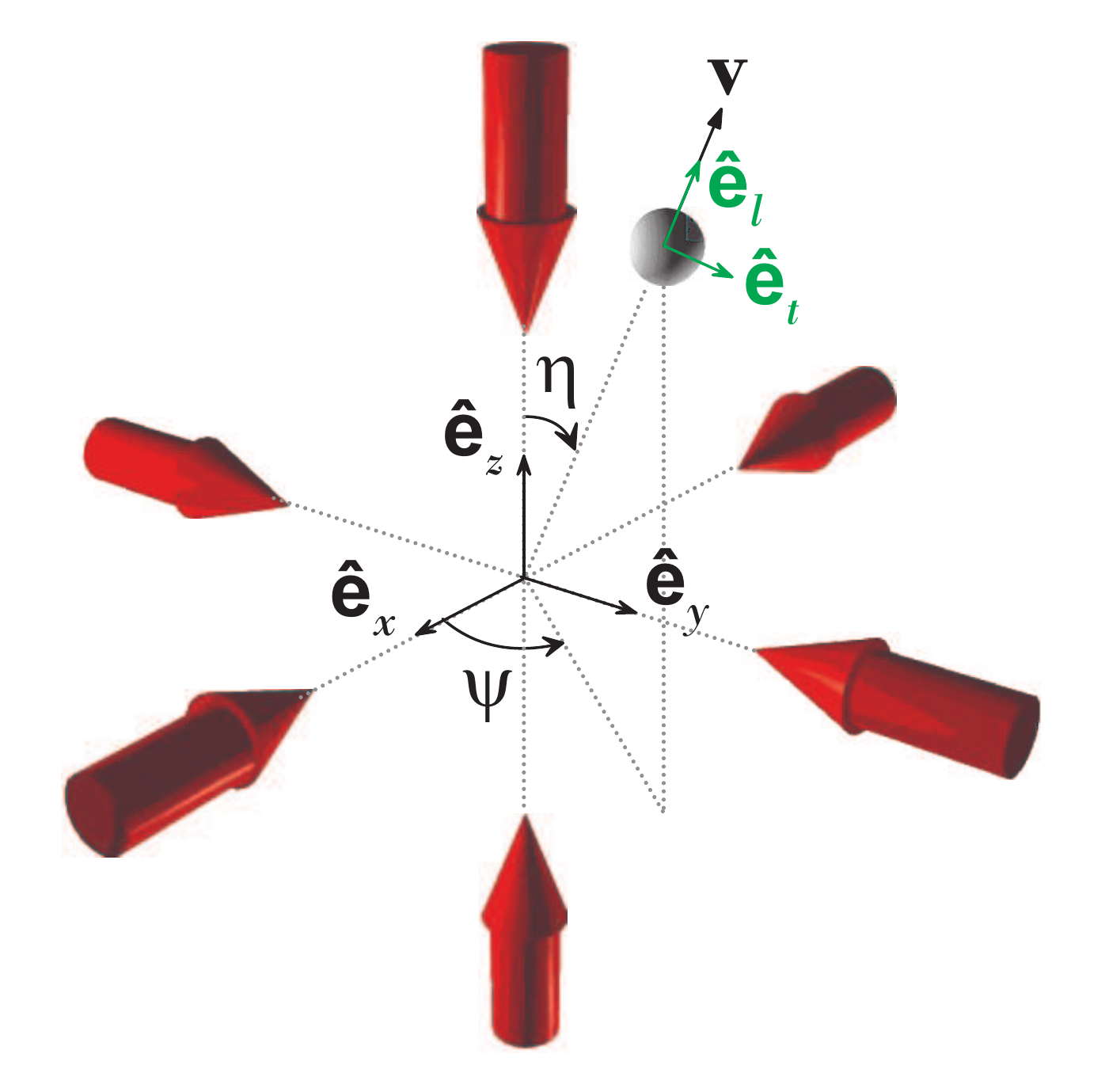}
    \caption{(Color online) 3D representation of the rectangular Cartesian coordinate system showing the conventions used in the whole paper. The cooling beams of the SBM configuration, that are supposed to be plane waves and thus having an infinite diameter, are here represented by red arrows. The direction followed by the atom relatively to the beams is defined by the angles $\eta$ and $\psi$. As we mainly consider the velocity space in this paper, the atom position is unimportant on this scheme and should be at the origin. Here, the atom is moved off-center for a better view. An example is given of vectors $\hat{e}_{l}$ and $\hat{e}_{t}$ along which the longitudinal and transverse components of the radiation-pressure-force are directed respectively.}
    \label{fig:Repere}
    \end{figure}

    Expression~(\ref{eq:Fcol1D}) gives a very good approximation of the real radiation pressure force undergone by the atoms in 1D optical molasses~\cite{Minogin1979}. However, this simple calculation does not take into account the coherent redistribution of photons from one wave to the other and/or any saturation effect due to the second wave. Thus, the generalization of the method used in Eq.~(\ref{eq:Fcol1D}) to a 3D configuration restricts its validity to very low saturation parameters ($s_{0}\ll 1$). Then, the average radiation-pressure forces exerted on an atom by three orthogonal standing waves (3D optical molasses) can be linearly added~\cite{Minogin1987,Lett1989}; the resulting force is:
    \begin{eqnarray}
    \begin{split}
    \bm{\mathrm{\mathcal{F}_{col}^{3D}}}=\bm{\mathrm{\mathcal{F}_{x}^{1D}}}+\bm{\mathrm{\mathcal{F}_{y}^{1D}}}+\bm{\mathrm{\mathcal{F}_{z}^{1D}}},
    \end{split}
    \label{eq:Fcol3D}
    \end{eqnarray}
    where the $x$, $y$, $z$ axis are defined by the three orthogonal standing-wave directions as shown in Fig.~\ref{fig:Repere}, drawn by vectors $\hat{e}_x$, $\hat{e}_y$ and $\hat{e}_z$ respectively.
    Here, the laser beams are supposed to be of equal intensities. We define a global saturation parameter for this cooling scheme, relative to the full light intensity, given by
    \begin{eqnarray}
    S_{col}=6\times s_{0}.
    \label{eq:Scol}
    \end{eqnarray}
    Each standing-wave set exerts a force on the atom, along the laser-beam direction. The resulting force of the three standing waves can be of any direction according to the direction of the atomic motion and its velocity. For the incoming comparison between different cooling configurations (see Section~\ref{sec:TheoreticalComparison}), it will be useful to express $\bm{\mathrm{\mathcal{F}_{col}^{3D}}}$ as a function of its longitudinal component $\mathcal{F}_{col}^{(l)}$, collinear to the atomic motion, and its transverse component $\mathcal{F}_{col}^{(t)}$, orthogonal to the atomic motion, drawn by vectors $\hat{e}_{l}$ and $\hat{e}_{t}$ respectively (see Fig.~\ref{fig:Repere}), as:
    \begin{eqnarray}
    \bm{\mathrm{\mathcal{F}_{col}^{3D}}}=\mathcal{F}_{col}^{(l)} \hat{e}_{l} + \mathcal{F}_{col}^{(t)} \hat{e}_{t}.
    \label{eq:Fcol3DLongPerp}
    \end{eqnarray}
    Indeed, we will show in section \ref{subsec:TransverseComponent} that, in the SBM scheme and at high atomic velocities, most of the force magnitude can be exerted in the plane transverse to the atomic motion.

    \subsection{\label{subsec:IsotropicLightConfiguration}Isotropic light configuration}

\begin{figure}[htb]
    \centering
    \includegraphics[width=0.9\linewidth]{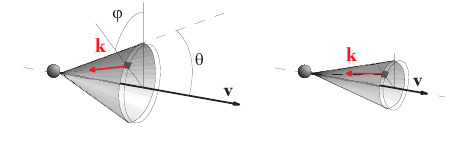}
    \caption{(Color online) Auto-adaptation of the resonance condition for an atom with velocity $\bm{\mathrm{v}}$ in presence of isotropic light. When the atom slows down due to the absorption of photons with wave vectors $\bm{\mathrm{k}}$, the cone fulfilling the Doppler resonance condition closes, being fully closed for $v=|\Delta|/k$.}
    \label{fig:ResonanceCone}
    \end{figure}

    In the isotropic light cooling scheme, we consider that the cavity into which the laser light is coupled generates an homogeneous light field of angular frequency $\omega$, where all the photons directions are evenly represented. $\omega$ is assumed to be red detuned. In order to fulfill the Doppler resonance condition given by Eq.~\ref{eq:ResonanceCondition}, an atom of resonance frequency $\omega_{0}$ will preferentially absorbs photons in the direction of angle $\theta$ such as $cos\theta=-\Delta/(kv)$, \emph{i.e.} belonging to a cone in 3D space as illustrated in Fig.~\ref{fig:ResonanceCone}.
    In the ILC configuration, for velocities $v\geq|\Delta|/k$, there are always photons to satisfy the resonance condition. As the atom slows down, the cone angle closes, being fully closed for $v=|\Delta|/k$. Then, considering the natural linewidth of the cooling transition, only some \emph{off-resonance} absorptions will occur.\\

    Let us consider that the light travelling in the solid angle $\delta\Omega$ around the wave vector $\bm{\mathrm{k}}_{\theta,\varphi}$ (such as $\|\bm{\mathrm{k}}_{\theta,\varphi}\|=k$, and whose direction is defined by the angles $\theta$ and $\varphi$ relatively to the atomic motion - see Fig.~\ref{fig:ResonanceCone}) acts as a single plane wave of intensity $\delta I$. Then, using expression (\ref{eq:F1Faisceau}) for low saturation parameters, we can express the elementary radiation-pressure force $\bm{\mathrm{\delta\mathcal{F}}}$ exerted by this wave as:
    \begin{eqnarray}
    \bm{\mathrm{\delta\mathcal{F}}}(\theta,\varphi)=\frac{\Gamma}{2} \delta s\hbar \bm{\mathrm{k}}_{\theta,\varphi}=\delta\mathcal{F}^{(l)} \hat{e}_{l}+\delta\mathcal{F}^{(t)} \hat{e}_{t},
    \end{eqnarray}
    where $\delta\mathcal{F}^{(l)}$ and $\delta\mathcal{F}^{(t)}$ are the longitudinal and transverse components of the elementary force, respectively. $\delta s=s_{\theta}\delta\Omega$ is the saturation parameter associated with the considered wave, where $s_{\theta}=\widetilde{s_{0}}/\left[1+4 \left(\Delta + k.v.cos\theta\right)^{2}/\Gamma^{2}\right]$. The total mean force $\bm{\mathrm{\mathcal{F}_{iso}}}$ exerted by an ILC scheme on an atom of velocity $\bm{\mathrm{v}}$, is then obtained by adding the elementary forces exerted in each space direction as:
    \begin{eqnarray}
    \bm{\mathrm{\mathcal{F}_{iso}}}= \iint \bm{\mathrm{\delta\mathcal{F}}}(\theta,\varphi).
    \label{eq:Fiso}
    \end{eqnarray}
    Due to rotational symmetry properties, the integration on $\varphi$ in Eq. (\ref{eq:Fiso}) eliminates the transverse component of the mean force, which leads to:
    \begin{eqnarray}
    \begin{split}
    \mathcal{F}_{iso} = \mathcal{F}_{iso}^{(l)} = &4\pi \widetilde{s_{0}} \times \hbar k \frac{\Gamma}{2} \left(\frac{\Gamma^{2}}{16 k^{2} v^{2}}\right)\\
    &\times \Bigg\{\ln \left[\frac{1+ 4\left(\Delta-k.v\right)^{2}/\Gamma^{2}}{1+ 4\left(\Delta+k.v\right)^{2}/\Gamma^{2}}\right]\\
    + \frac{4\Delta}{\Gamma} \Bigg(\arctan &\left[\frac{2(\Delta+k.v)}{\Gamma}\right] - \arctan \left[\frac{2(\Delta-k.v)}{\Gamma}\right]\Bigg)\Bigg\}.
    \label{eq:FisoL}
    \end{split}
    \end{eqnarray}
    As previously mentioned in Section~\ref{subsec:6BeamMolassesConfiguration}, adding the elementary forces as in Eq.~(\ref{eq:Fiso}) will restrict the use of expression (\ref{eq:FisoL}) to very low saturation parameters. Since the transverse component of $\bm{\mathrm{\mathcal{F}_{iso}}}$ is null, the ILC scheme allows a deceleration that is always anti-parallel to the atomic motion, whatever its direction, as opposed to other cooling schemes.\\

    For the comparisons to come, and as done previously for the SBM configuration (see Eq.~(\ref{eq:Scol})), we need to define a global saturation parameter for the ILC scheme. The saturation parameter at resonance $\widetilde{s_{0}}$ introduced previously for a wave of direction given by angles $\theta$ and $\varphi$, can also be written as $\widetilde{s_{0}}=d S_{iso}/d\Omega$, where $d S_{iso}$ is the global saturation parameter at resonance for waves contained in the solid angle $d\Omega$. The cooling field isotropy then leads to:
    \begin{eqnarray}
    S_{iso}=4\pi \widetilde{s_{0}}.
    \end{eqnarray}

\section{\label{sec:TheoreticalComparison}Theoretical comparison of the two cooling configurations}

We now compare both configurations considering separately the component of the radiation-pressure force collinear to the atom's velocity (responsible of the cooling) and the transverse component. Here, we mainly study the spacial properties of the radiation-pressure force, \textit{i.e.} the force magnitude as a function of the direction of the atomic motion. These 3D comparisons also bring us to 1D comparisons, in which we are lead to consider mean values of the force components in the SBM case, averaged over all possible atomic incidences relatively to the beams. The relative capture efficiency of the two cooling schemes is then discussed. The computations are performed for the $D_{2}$ line of a Cs atom (\emph{i.e.} $\lambda= 852.35$~nm and $\Gamma/2\pi=5.22$~MHz), for a red detuning $\Delta=-2\Gamma$ and saturation parameters $S_{iso}=S_{col}=0.1$.

    \subsection{\label{subsec:LongitudinalComponent}Longitudinal component}

\begin{figure}[htb]
    \centering
    \includegraphics[trim=10mm 10mm 20mm 10mm, width=\linewidth]{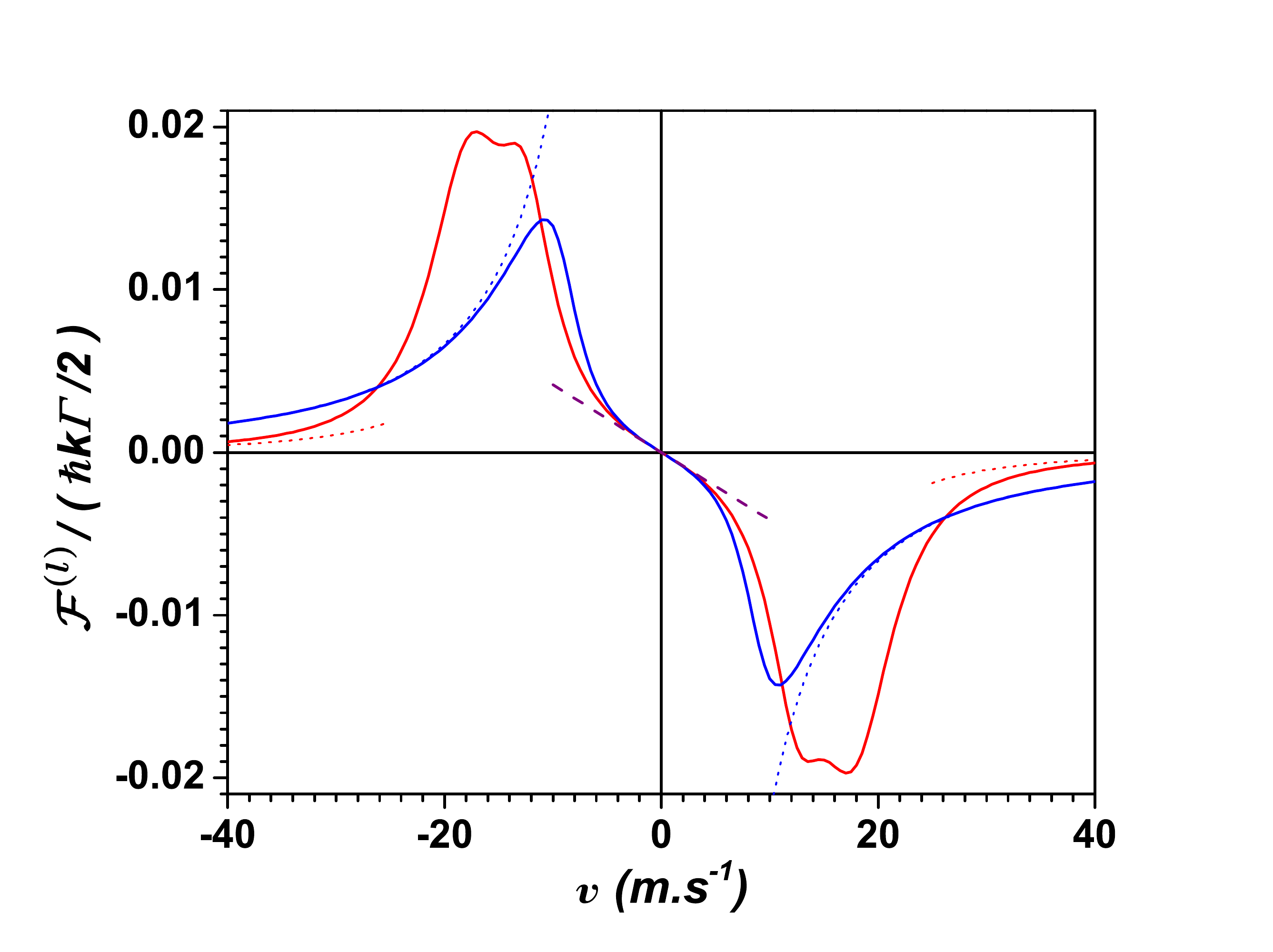}
    \caption{(Color online) Magnitude of the radiation-pressure-force longitudinal component exerted along an arbitrary direction (here $\eta=\psi=\nicefrac{\pi}{4}$ - see Fig~\ref{fig:Repere}) vs atomic velocity, in a SBM configuration (red solid line) and in an ILC configuration (blue solid line). The linear approximation of both forces at low velocity given by Eq.~\ref{eq:FlongFaiblesVitesses} is shown with a purple dashed line, while at high velocity, the Taylor series expansions of $\mathcal{F}_{col}^{(l)}$ (Eq.~\ref{eq:FcolLGrandesVitesses}) and $\mathcal{F}_{iso}^{(l)}$ (Eq.~\ref{eq:FisoGrandesVitesses}) are plotted with dotted red and blue lines respectively. The calculations are performed for a detuning $\Delta=-2\Gamma$ and saturation parameters $S_{col}=S_{iso}=0.1$.}
    \label{fig:Flong1DComparison}
    \end{figure}

    We first proceed to a 1D comparison of the two cooling configurations, providing in Fig.~\ref{fig:Flong1DComparison} the variations of the radiation-pressure-force longitudinal components defined in Eq. \ref{eq:Fcol3DLongPerp} and \ref{eq:FisoL}, as a function of the velocity, for an atom moving along an arbitrary space direction.
    In order to point out general properties, we restrain this 1D comparison to both extrema, i.e. at low and high velocity only. Making relevant comparisons in the intermediate velocity range requires a 3D point of view that we give in the next paragraph. One can first observe that, at low velocity, both forces merge. Indeed, for $|k v|\ll|\Delta|$ and $|k v|\ll\Gamma$, $\mathcal{F}_{col}^{(l)}$ and $\mathcal{F}_{iso}^{(l)}$ can be reduced to the same linear approximation:
    \begin{eqnarray}
    \begin{split}
    \mathcal{F}^{(l)}\sim -&\alpha v \hspace{0.3cm}\\
    \text{with} \hspace{0.3cm} &\alpha=-4\hbar k^{2}\left(\frac{S}{6}\right)\times \frac{\left(2\Delta/\Gamma\right)}{\left[1+\left(2\Delta/\Gamma\right)^{2}\right]^{2}},
    \label{eq:FlongFaiblesVitesses}
    \end{split}
    \end{eqnarray}
    where $S$ must be replaced by the appropriate global saturation parameter $S_{col}$ or $S_{iso}$. On the other hand, at high velocity, the two forces behave differently, $\mathcal{F}_{col}^{(l)}$ decreasing faster than $\mathcal{F}_{iso}^{(l)}$. This is confirmed by the following Taylor series expansions of both forces at $v\rightarrow \infty$ (only valid when $\eta$ and $\psi$ differ from $0$ mod $\nicefrac{\pi}{2}$ in case of Eq.~\ref{eq:FcolLGrandesVitesses}):
    \begin{eqnarray}
    \begin{split}
    \mathcal{F}_{col}^{(l)}=6 s_{0}\times \hbar k \frac{\Gamma}{2} &\times \Bigg[\frac{\Gamma^{2}\Delta}{6k^{3}}\Bigg(\frac{4}{sin^{2}\eta\times sin^{2}(2\psi)}\\
    &+\frac{1}{cos^{2}\eta}\Bigg)\Bigg]\times\frac{1}{v^{3}}+\mathcal{O}\left(\frac{1}{v^{5}}\right),
    \label{eq:FcolLGrandesVitesses}
    \end{split}
    \end{eqnarray}
    \begin{eqnarray}
    \begin{split}
    \mathcal{F}_{iso}^{(l)}=4\pi \widetilde{s_{0}}\times \hbar k \frac{\Gamma}{2} &\times \Bigg[\frac{v}{|v|}\left(\frac{\pi\Gamma\Delta}{4k^{2}}\right)\times\frac{1}{v^{2}}\\ &-\left(\frac{\Gamma^{2}\Delta}{2k^{3}}\right)\times\frac{1}{v^{3}}\Bigg]+\mathcal{O}\left(\frac{1}{v^{5}}\right),
    \label{eq:FisoGrandesVitesses}
    \end{split}
    \end{eqnarray}
    the first term of $\mathcal{F}_{col}^{(l)}$ being in $v^{-3}$, while it is in $v^{-2}$ for $\mathcal{F}_{iso}^{(l)}$. This slower decrease at high velocity of the radiation-pressure-force longitudinal component in the ILC scheme can be explained by referring to Fig.~\ref{fig:ResonanceCone}. Indeed, in such a configuration and as previously mentioned in section~\ref{subsec:IsotropicLightConfiguration}, there are always photons satisfying the resonance condition given by Eq.~\ref{eq:ResonanceCondition} for atoms with velocities $|v|\geq |\Delta|/k$ (whatever the direction of the atomic motion). However, in the SBM configuration, for a given velocity value, there will be only few space directions along which the moving atom will encounter resonant photons, these directions being the ones for which the cone described in Fig.~\ref{fig:ResonanceCone} coincides with at least one beam. In other words, for a given space direction of a SBM configuration (as in Fig.~\ref{fig:Flong1DComparison}) and for velocities such as $|v|>|\Delta|/(k.cos\theta)$, atoms will absorb light only \emph{off resonantly}. Actually, this is this difference between the two cooling schemes that made the authors of ref.~\cite{Ketterle1992} suggest that ILC could be more effective than the usual SBM configuration, since the capture range should be extended in the ILC case.\\

\begin{figure*}
    \centering
    \includegraphics[width=0.35\linewidth]{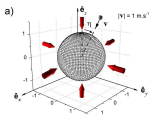}
    \includegraphics[width=0.35\linewidth]{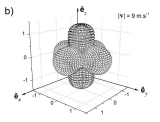}\\
    \includegraphics[width=0.35\linewidth]{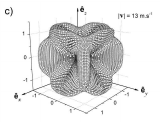}
    \includegraphics[width=0.35\linewidth]{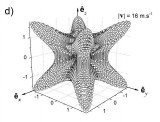}\\
    \includegraphics[width=0.35\linewidth]{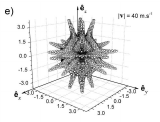}
    \includegraphics[width=0.35\linewidth]{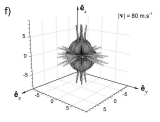}
    \caption{(Color online) Magnitude of the radiation-pressure-force longitudinal component exerted by a SBM configuration on an atom with velocity $\bm{\mathrm{v}}$, as a function of the atomic incidence relatively to the six beams. The force has been normalized to the one calculated in isotropic light for the same cooling parameters ($\mathcal{F}_{col}^{(l)}/\mathcal{F}_{iso}^{(l)}$). Cooling beams, respectively directed along $\pm x$, $\pm y$ et $\pm z$, are represented by red arrows in figure $(a)$. The calculations are performed for an optical detuning $\Delta = -2\Gamma$, saturation parameters $S_{col}=S_{iso}=0.1$ and the values of $|\bm{\mathrm{v}}|$ indicated on each graph.}
    \label{fig:Flong3DComparison}
    \end{figure*}

    The following paragraph is dedicated to a 3D comparison of the two cooling configurations. First of all, it has to be mentioned that in the ILC configuration the magnitude of the radiation-pressure force $\mathcal{F}_{iso}^{(l)}$, is independent of the atom direction. A 3D representation of the force magnitude as a function of the direction would lead to observe a perfect sphere. We show in Fig.~\ref{fig:Flong3DComparison} the magnitude of the longitudinal component of the radiation-pressure force exerted by a six-beam molasses on an atom of velocity $v$, as a function of the atomic incidence relatively to the cooling beams.
    On these graphs, the distance from a point to the origin gives the magnitude of the force component for an atom following the direction defined by this point and the origin. The graphs are computed for different values of the atomic velocity $v$ ranging from $1$ to $80$~m.s$^{-1}$. In order to compare the two cooling schemes, the magnitude of the force component is normalized to the magnitude of the isotropic-light force $\mathcal{F}_{iso}^{(l)}$. For low-enough velocities such as $|kv| \ll |\Delta|$ (\emph{i.e.} $v\lesssim 0.1\times 2\Gamma/k \sim 1$~m.s$^{-1}$), and according to the resonance condition of Eq.~(\ref{eq:ResonanceCondition}), the magnitude of the longitudinal force is almost independent of the angle $\theta$ between the atom's motion and the light field direction. The force calculated in the SBM case is thus equal to the one calculated for an ILC configuration in the same conditions (see Eq.~\ref{eq:FlongFaiblesVitesses}). That is why we observe a sphere with a unit radius in Fig.~\ref{fig:Flong3DComparison}-a. Then, for $v \sim |\Delta|/k$ ($\sim 9$~m.s$^{-1}$), the aperture of the cone of resonance shown in Fig.~\ref{fig:ResonanceCone} being almost null, the atoms mainly absorb photons having the same direction as $\bm{\mathrm{v}}$. This leads to a maximum of the longitudinal force along the beams axes of the SBM configuration, as observed in Fig.~\ref{fig:Flong3DComparison}-b. For $v=13$~m.s$^{-1}$ (see Fig.~\ref{fig:Flong3DComparison}-c), the resonance condition is fulfilled when $\theta\sim \nicefrac{\pi}{4}$, which leads to a force maximum along the diagonals of the planes defined by two pairs of beams. Indeed, in this case, when following the $(1,1,0)$ direction for example, the atoms can \emph{resonantly} absorb photons from two orthogonal beams. For $v=16$~m.s$^{-1}$ (see Fig.~\ref{fig:Flong3DComparison}-d), the resonance condition is fulfilled for $cos \theta\sim\nicefrac{1}{\sqrt{3}}$. Hence, the maximum amplitude of $\mathcal{F}_{col}^{(l)}$ will be reached along the system's diagonals, allowing the atoms to \emph{resonantly} absorb photons from three orthogonal beams, as along the $(1,1,1)$ direction for example. At first glance, the last two graphs of Fig.~\ref{fig:Flong3DComparison} seem harder to interpret due to the many occurring resonances. When looking carefully to the graphs \ref{fig:Flong3DComparison}-e and \ref{fig:Flong3DComparison}-f, we can distinguish four maxima around each beam direction, \emph{i.e.} $24$ in total, these maxima getting closer to the beams axes when the velocity is increased from $40$~m.s$^{-1}$ to $80$~m.s$^{-1}$. This can be interpreted by pursuing the previous reasoning. Indeed, if we now consider the limit $v\rightarrow \infty$, Eq.~(\ref{eq:ResonanceCondition}) leads to $\theta\rightarrow\pi/2$, which means that the cone of resonance is fully opened. In that case, an atom moving along the beams axes should be able to \emph{resonantly} absorb photons from up to four different beams. But meanwhile, the magnitude of the longitudinal force tends to $0$ and makes this case unreachable in theory. Nevertheless, the maxima observed in graphs \ref{fig:Flong3DComparison}-e and \ref{fig:Flong3DComparison}-f are effectively due to the absorption of photons in four different beams, but \emph{off resonantly}. Indeed, as pointed out in Section~\ref{subsec:IsotropicLightConfiguration}, the resonance condition is not as restrictive as given by Eq.~(\ref{eq:ResonanceCondition}). Actually, the cone shown in Fig.~\ref{fig:ResonanceCone} has a non-null thickness given by a Lorentzian distribution law with a FWHM equal to $\Gamma$, allowing atoms moving along directions slightly shifted compared to the ideal cases described so far, to fulfill the resonance condition with one, two, three or four beams simultaneously.\\

\begin{figure}[tb]
    \centering
    \includegraphics[trim=10mm 10mm 20mm 10mm, width=\linewidth]{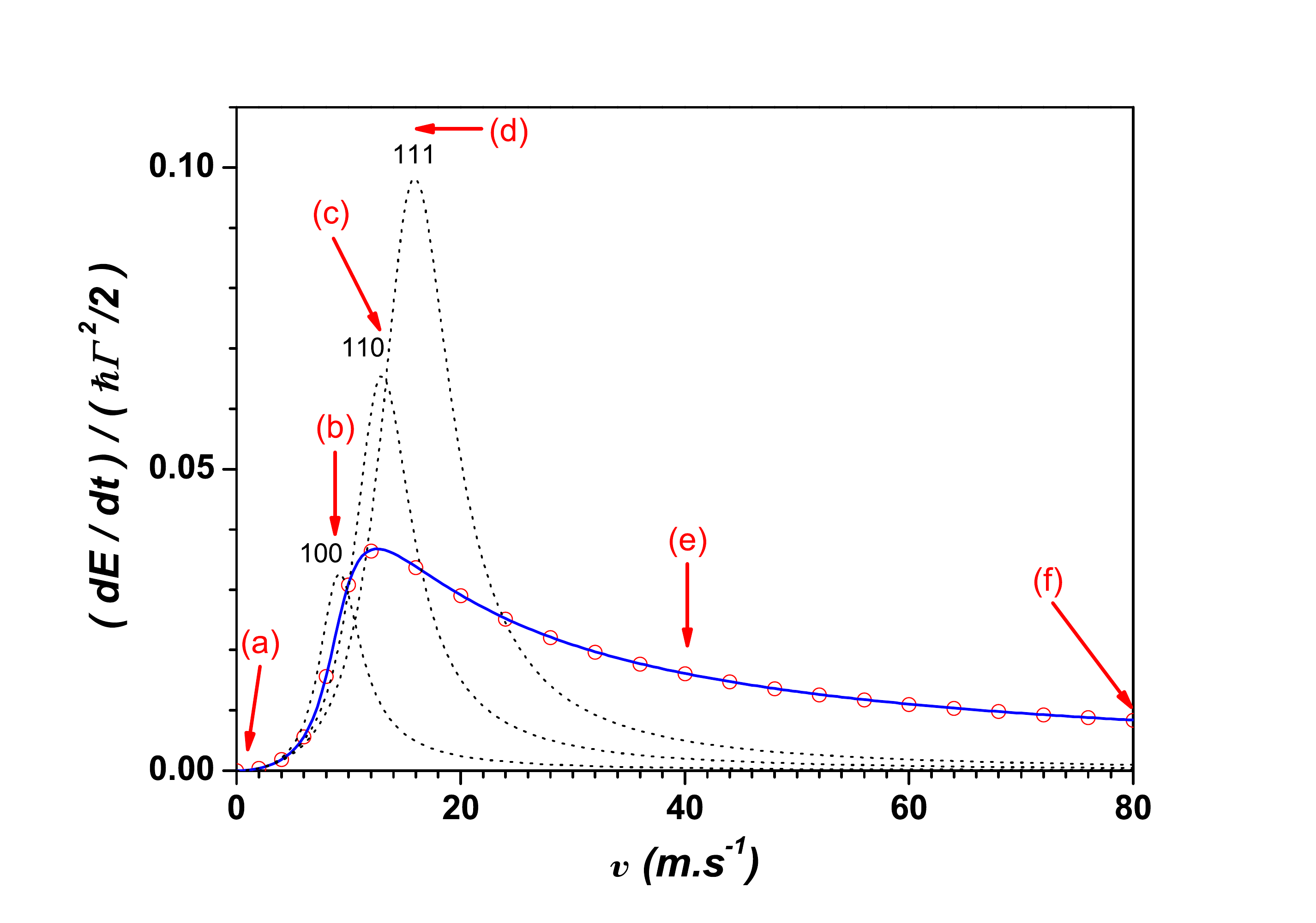}
    \caption{(Color online) Cooling rate vs atomic velocity in an ILC configuration ($\bm{\mathrm{\mathcal{F}_{iso}}}\cdot \bm{\mathrm{v}}$ - blue solid line), and in a SBM configuration ($\bm{\mathrm{\mathcal{F}_{col}^{3D}}}\cdot\bm{\mathrm{v}}$ - gray dotted lines) with laser beams along the $\pm x$, $\pm y$, $\pm z$ directions and $\bm{\mathrm{v}}$ along the $(1,0,0)$, $(1,1,0)$ and $(1,1,1)$ directions. The mean cooling rate $\overline{\bm{\mathrm{\mathcal{F}_{col}^{3D}}}\cdot\bm{\mathrm{v}}}$ has also been plotted $(\circ)$ - see Eq.~\ref{eq:CollimatedLightMeanCoolingRate}. The calculations are performed for a detuning $\Delta = -2\Gamma$ and saturation parameters $S_{col}=S_{iso}=0.1$. The indicators $(a)$ to $(f)$ spot the velocity values for which the graphs of Fig.~\ref{fig:Flong3DComparison} have been calculated.}
    \label{fig:CoolingRates}
    \end{figure}

    We now compare both configurations in term of cooling rates, \emph{i.e.} regarding the energy loss per atom and per unit of time given by $dE(v)/dt=\bm{\mathrm{\mathcal{F}(v)}}\cdot\bm{\mathrm{v}}$. In the case of the SBM configuration, we have first computed the cooling rates variations as functions of the atomic velocity for the three specific directions $(1,0,0)$, $(1,1,0)$ and $(1,1,1)$, see Fig.~\ref{fig:CoolingRates}.
    Such a comparison has already been made by Ketterle \textit{et al.}~\cite{Ketterle1992}. Assuming that, for velocities $v>|\Delta|/k$, the most favorable direction lied symmetrically between the three pairs of laser beams of a SBM (\textit{i.e.} the $(1,1,1)$ direction for example), authors of Ref.~\cite{Ketterle1992} concluded that the ILC scheme had a much faster cooling rate than SBM at high velocity, resulting in a larger velocity capture range, as can be observed in Fig.~\ref{fig:CoolingRates} for $v\gtrsim 24$~m.s$^{-1}$. As a consequence, isotropic light was expected to be more efficient at cooling an atomic vapor than the SBM configuration. Nevertheless, these conclusions were too fast. Actually, the graphs of Fig.~\ref{fig:Flong3DComparison} point out that the most favorable direction only lies symmetrically between the three pairs of laser beams of a SBM for a given value of the atomic velocity (see Fig.~\ref{fig:Flong3DComparison}-d). In other words, the radiation-pressure force reaches a maximum for some atomic directions relatively to the cooling beams, but these favored directions depend on the atomic velocity $v$, the optical detuning $\Delta$, and the saturation parameter $S_{col}$. These directions are not always the particular cases $(1,0,0)$, $(1,1,0)$ and $(1,1,1)$ (see Fig.~\ref{fig:Flong3DComparison}-e and \ref{fig:Flong3DComparison}-f).\\

    In the case of a background vapor and a SBM configuration, where all atomic incidences relatively to the six beams are met with the same probability, we find more relevant to evaluate a mean cooling rate, \emph{i.e.} a cooling rate averaged over all space directions of the atomic motion, to be compared with the ILC case. This mean cooling rate, reported in Fig.~\ref{fig:CoolingRates}, is given by:
    \begin{eqnarray}
    \overline{\bm{\mathrm{\mathcal{F}_{col}^{3D}}}\cdot\bm{\mathrm{v}}} =\frac{v}{4\pi} \int\limits_{0}\limits^{2\pi}\int\limits_{0}\limits^{\pi} \mathcal{F}_{col}^{(l)} \hspace{0.1cm} sin\eta d\eta d\psi,
    \label{eq:CollimatedLightMeanCoolingRate}
    \end{eqnarray}
    where $\eta$ and $\psi$ are the polar and azimuthal angles respectively, defining the direction followed by the atom in the spherical coordinate system (see Fig.~\ref{fig:Repere}). We can observe that this mean cooling rate is not only higher than the one calculated for the $(1,1,1)$ direction for $v\gtrsim 24$~m.s$^{-1}$, but is also strictly equal to the cooling rate calculated for the ILC configuration on the full range of velocities between $0$ and $80$~m.s$^{-1}$. This result is not as surprising as it could seem. Indeed, the mean cooling rate given by Eq.~\ref{eq:CollimatedLightMeanCoolingRate} in the case of a SBM configuration could also be written as six times the mean cooling rate evaluated for one beam only. In such a calculation, the reference direction is given by the beam itself, and the average is performed on all possible atomic incidences relatively to this beam. In a reverse way, when evaluating the cooling rate in the ILC case, the atomic direction is taken as the reference direction, and the average is performed on all possible wave incidences (see Eq.~\ref{eq:Fiso}). A straightforward calculation finally shows that both cooling rates are equal when $\tilde{s_0}=6 s_0/4\pi$, \emph{i.e.} when $S_{iso}=S_{col}$.

    \subsection{\label{subsec:TransverseComponent}Transverse component}

\begin{figure*}
    \centering
    \includegraphics[width=0.35\linewidth]{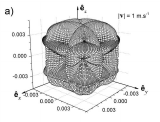}
    \includegraphics[width=0.35\linewidth]{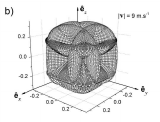}\\
    \includegraphics[width=0.35\linewidth]{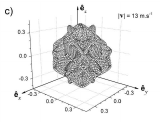}
    \includegraphics[width=0.35\linewidth]{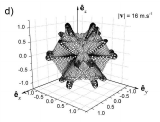}\\
    \includegraphics[width=0.35\linewidth]{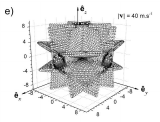}
    \includegraphics[width=0.35\linewidth]{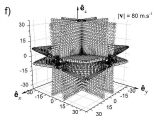}
    \caption{Magnitude of the radiation-pressure-force transverse component exerted by a SBM configuration on an atom with velocity $\bm{\mathrm{v}}$, as a function of the atomic incidence relatively to the six beams. As in Fig.~\ref{fig:Flong3DComparison}, the force has been normalized to the isotropic-light force ($\mathcal{F}_{col}^{(t)}/\mathcal{F}_{iso}^{(l)}$). The calculations are performed for an optical detuning $\Delta = -2\Gamma$, saturation parameters $S_{col}=S_{iso}=0.1$ and the values of $|\bm{\mathrm{v}}|$ indicated on each graph.}
    \label{fig:Ftrans3DComparison}
    \end{figure*}

\begin{figure}[tb]
    \centering
    \includegraphics[trim=10mm 10mm 20mm 10mm, width=\linewidth]{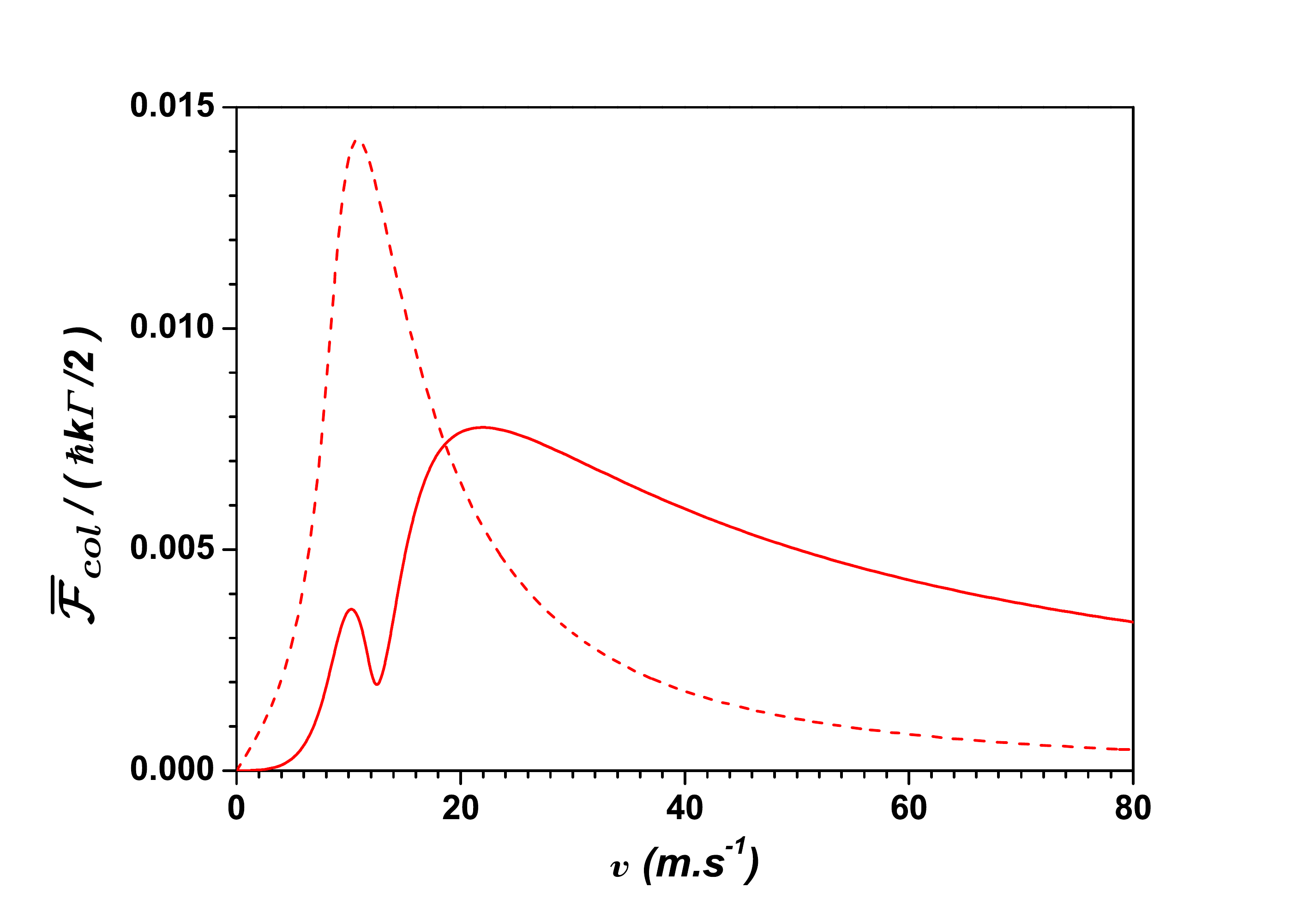}
    \caption{(Color online) Mean value of the radiation-pressure-force transverse component ($\overline{\mathcal{F}_{col}^{(t)}}$ - red solid line) vs atomic velocity in a SBM configuration. The mean value of the longitudinal component ($\overline{\mathcal{F}_{col}^{(l)}}$ - red dashed line) has also been reported for reference. The calculations are performed for an optical detuning $\Delta = -2\Gamma$ and a saturation parameter $S_{col}=0.1$.}
    \label{fig:SBMFtransMoy}
    \end{figure}

    Fig.~\ref{fig:Ftrans3DComparison} shows the 3D variations of the magnitude of the radiation-pressure-force transverse component exerted by a six-beam molasses on an atom of velocity $v$, in the same conditions as in Fig.~\ref{fig:Flong3DComparison}.
    We recall that, without consideration of the recoil induced by spontaneous emissions, there is no transverse force in the ILC configuration for all atomic velocities. To compare the force magnitudes, here also the magnitude of the transverse force calculated for a SBM configuration is normalized to the isotropic-light force $\mathcal{F}_{iso}^{(l)}$. In order to interpret the graphs of Fig.~\ref{fig:Ftrans3DComparison}, we can first point out the specific cases for which the force transverse component $\mathcal{F}_{col}^{(t)}$ is null in a SBM. For an atom moving along an arbitrary direction, the transverse force is directed along a given direction. But if we consider an atom moving along one of the rotational symmetry axis of the system, and if we apply a rotation that leaves the system unchanged, then the direction of the force transverse component should rotate accordingly. The only force that can be directed in more than one direction being obviously null, we deduce that $\mathcal{F}_{col}^{(t)}=0$ along all the rotational symmetry axis (of $2^{\textrm{nd}}$, $3^{\textrm{rd}}$ and $4^{\textrm{th}}$ order) of the system. For example, this can be guessed along the $(1,0,0)$, $(1,1,0)$ and $(1,1,1)$ directions on most of the graphs of Fig.~\ref{fig:Ftrans3DComparison}. If we now compare the graphs of Fig.~\ref{fig:Flong3DComparison} and \ref{fig:Ftrans3DComparison} for increasing atomic velocities, we first note that for low-enough velocities such as $|k.v| \ll |\Delta|$ (see Fig.~\ref{fig:Flong3DComparison}-a and \ref{fig:Ftrans3DComparison}-a while noting the difference in scale), the force transverse component is very weak in comparison to the longitudinal component. Indeed, in this case, the radiation-pressure forces exerted by the six beams on an atom being almost independent of the incidence angle $\theta$ (see Fig.~\ref{fig:ResonanceCone}), and thus having about the same amplitude, they almost compensate each other in the plane transverse to the atomic motion. On the other hand, for high velocities (see Fig.~\ref{fig:Flong3DComparison}-f and \ref{fig:Ftrans3DComparison}-f for example) and for particular directions, one can note that most of the radiation-pressure force is exerted in the plane transverse to the atomic motion. This is expected, since the higher is $v$, the larger is the top angle $\theta$ of the cone to which belong the photons fulfilling the resonance condition given by Eq.~\ref{eq:ResonanceCondition} (see Fig.~\ref{fig:ResonanceCone}). This can also be observed when comparing the variations of $\overline{\mathcal{F}_{col}^{(l)}}$ and $\overline{\mathcal{F}_{col}^{(t)}}$ for increasing velocities (see Fig.~\ref{fig:SBMFtransMoy}), where we remind that the upper bar on the radiation-pressure-force components denotes an average over all possible atomic incidences relatively to the six beams.
    \\

\begin{figure}[tb]
    \centering
    \includegraphics[trim=10mm 10mm 20mm 10mm, width=\linewidth]{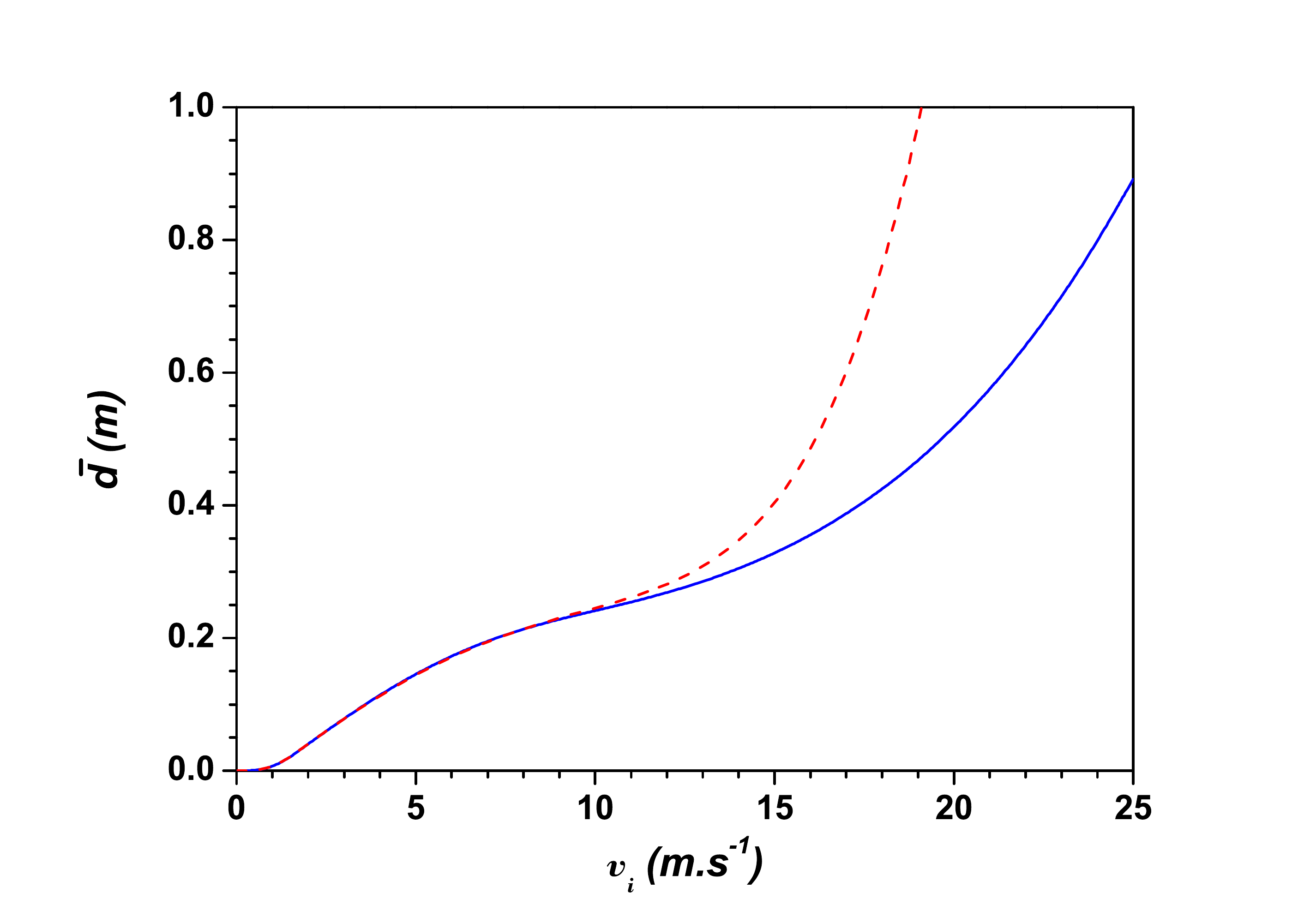}
    \caption{(Color online) Mean distance needed for an atom with initial velocity $v_{i}$ to be decelerated down to $v_{f}=1$~m.s$^{-1}$, when undergoing the radiation-pressure forces exerted in an ILC configuration (blue solid line) or in a SBM (red dashed line). The calculations are performed for an optical detuning $\Delta = -2\Gamma$ and saturation parameters $S_{col}=S_{iso}=0.1$.}
    \label{fig:MeanCaptureDistance}
    \end{figure}

    From Section~\ref{subsec:LongitudinalComponent} and the previous paragraph, we now understand that it is not possible to restrict the comparison between the ILC configuration and the SBM to a given direction of the atomic incidence, or even to the longitudinal component of the radiation-pressure force. On the one hand, we have shown that, for a six-beam molasses configuration, the direction along which the radiation-pressure force reaches a maximum fully depends of the considered value of the atomic velocity (see Fig.~\ref{fig:Flong3DComparison}). On the other hand, we have highlighted what mainly differentiates the two cooling configurations : the transverse component of the radiation-pressure force (see Fig.~\ref{fig:Ftrans3DComparison}). Indeed, if the force transverse component is always null in isotropic light, it can become the main component of the force exerted in a six-beam molasses for atoms with high velocities, as observed in Fig.~\ref{fig:SBMFtransMoy} for $v>18$~m.s$^{-1}$. In the SBM case, the existence of this transverse component will lead to an increase of the atomic-trajectories length during the cooling process, when compared to the ILC case. This specificity of the SBM configuration will limit, for a given capture volume, the Doppler capture velocity to a lower value than the one expected in the ILC case. This is illustrated in Fig.~\ref{fig:MeanCaptureDistance}, where we have reported the distance $\bar{d}$ needed for an atom with initial velocity $v_{i}$ to be decelerated down to $v_{f}=1$~m.s$^{-1}$, for both cooling configurations.
    The computations are performed by solving the equations of motion for an atom undergoing the radiation-pressure forces given by Eq.~\ref{eq:Fcol3D} and \ref{eq:FisoL} respectively. $v_{f}=1$~m.s$^{-1}$ has been chosen as a typical value of the capture velocity for the sub-Doppler cooling processes~\cite{Dalibard1989} (not taken into account in this paper). In the case of the SBM configuration, the distance $d$ has been averaged over all possible directions of the vector $\bm{\mathrm{v_{i}}}$. Due to the low saturation parameters used for our calculations, the distances obtained here are significantly higher than the ones observed in conventional cooling experiments (for the same velocity range as considered in Fig.~\ref{fig:MeanCaptureDistance} and saturation parameters on the order of $10$, $\bar{d}$ is usually of few centimeters). Nevertheless, Fig.~\ref{fig:MeanCaptureDistance} clearly shows that if both schemes are equivalent at low velocity (here below $7$~m.s$^{-1}$), the distance needed to decelerate the atoms from $v_{i}$ to $v_{f}$ increases faster at high velocity in the SBM case. Therefore, less atoms should be cooled from the low tail of the Maxwell-Boltzmann distribution, when compared to the ILC case.\\

    Here, we can finally reveal an interesting feature from Fig.~\ref{fig:Flong3DComparison} and \ref{fig:Ftrans3DComparison}. Indeed, if we consider the load of a 3D optical molasses with a 2D Magneto-Optical-Trap (MOT), the direction of injection of the cold atom beam would not matter in case of use of an ILC scheme. On the other hand, it seems that it should matter for a SBM configuration. Thus, for given values of the saturation parameter and detuning of the molasses beams, and for a given mean velocity of the cold atom beam, the 2D MOT should be oriented in a specific direction relatively to the molasses beams in order to maximize the longitudinal component of the cooling force, while minimizing its transverse component (\emph{i.e.} preferentially along one of the rotational symmetry axis of the configuration).

\section{\label{sec:level1}Conclusion}

We have recalled the expressions of the light-pressure forces at low saturation in a six-beam molasses, and in an isotropic light cooling configuration. The spatial properties of the forces, \textit{i.e.} their magnitude with respect to the direction of the atomic motion, have been investigated in the case of a thermal vapor. Naturally, in the ILC case the longitudinal force (or the related cooling rate) is direction independent. In SBM, on the contrary, there are directions where the force (or cooling rate) can be larger or smaller than in isotropic light. These directions depend on the atomic velocity magnitude. Nevertheless, assuming that cooling rates averaged over all directions are more representative in the case of a background gas, we have shown that the mean cooling rates are equal in both cases. A maximum is reached for velocity $v\sim|\Delta|/k$, followed by a slow decrease for increasing velocities. On the contrary, the two cooling schemes highly differ when considering the force component transverse to the movement. There is no transverse force for an ILC configuration (let's remember that, in this paper, we do not take into account the transverse momentum spreading induced neither by spontaneous emission nor by stimulated emission~\cite{Aardema1996}). In SBM, a transverse force exists, direction and velocity dependent. When averaged over all possible atomic incidences relatively to the six beams, the related mean value of this transverse force first mainly increases with $v$, becoming larger than the longitudinal component of the force for velocities higher than $v\sim 2|\Delta|/k$. Follows a decrease which is slower than the one observed for the longitudinal component, leading to a force exerted essentially in the plane transverse to the atomic motion at high velocity. Finally, we have computed the mean distance needed to slow an atom from the background vapor close to sub-Doppler capture velocities ($\sim 1$~m.s$^{-1}$) in both configurations, pointing out that the force transverse component met in the SBM case should lead to observe smaller Doppler capture velocities than in the ILC scheme.\\

Although the ILC configuration tolerates only pretty small apertures in the diffusive (or reflective) cavity around the vapor cell, which can restrict its applications, it has several advantages. It can be easier to implement on a 3D cooling experiment than the SBM configuration; it is expected to be more robust (not subject to beam misalignments or intensity imbalances), and it should also be less power-consuming owing to the recycling of cooling light by the cavity. As in Ref.~\cite{Wang1994}, we have considered in this paper that the light field generated by the cavity used in the ILC scheme was isotropic and homogeneous at a macroscopic scale, \emph{i.e.} on the whole cavity volume. Only intensity inhomogeneities along the tube axis have been considered in Ref.~\cite{Ketterle1992,Batelaan1994,Aardema1996}, ignoring the existence of a peaked intensity profile along the tube (or sphere) radius~\cite{Guillemot1999,Pottie2003}. The existence of such intensity inhomogeneities in the ILC configuration deeply affects the cooling performances that can be expected, and will be discussed in further studies.\\

It would be interesting to bring the theoretical comparison presented in this paper to the domain of high saturation parameters (such as $S_{col}=S_{iso}=10$ for example, value that is more commonly met in practice), to see if the expected superiority of ILC vs SBM still holds. The diffusion of the atom momentum should then be considered, since it has been shown that extra transverse diffusion is expected in the ILC configuration due to stimulated emission~\cite{Aardema1996}. Moreover, the highest probability for multiphoton processes in these conditions, would imply to consider multilevel atoms with at least three Zeeman sub-levels for the ground state, and five Zeeman sub-levels for the exited state. Indeed, if the Doppleron resonances~\cite{Kyrola1977} that occur at high intensity (deeply altering the shape of the radiation-pressure force vs atomic velocity) can even be met in the two-level atom model~\cite{Minogin1979}, they start to compete with even-order multiphoton transitions (responsible of sub-Doppler cooling) from the $(3+5)$-level atom model~\cite{Chang2002}. Even if the cooling field is assumed to be homogeneous at a macroscopic level for both configurations, the field properties at the microscopic scale would have to be considered for the calculations. These field properties would then depend, on the one hand, on the phase difference between the cooling beams for a SBM configuration, and on the other hand, on the roughness of the cavity inner surface that generates a speckle field in the ILC case. However, considering a multilevel atom moving in such laser fields with multidimensional periodicity~\cite{Molmer1991} will make the task difficult.

\section*{Acknowledgements}
 We would like to thank Natascia Castagna for helpful discussions. S. Trémine would like to thank Noël Dimarcq for having welcomed him on the HORACE project and thus having made this research possible. This work was financially supported by the French space agency (CNES). The research of S. Trémine has been made possible by a fellowship of the CNES and EADS-Sodern.


\end{document}